\RequirePackage{lineno}

\documentclass[a4,11pt]{article}

\usepackage[utf8]{inputenc}
\usepackage[T1]{fontenc}
\usepackage{bm}

\usepackage{authblk}

\usepackage{hyperref} 
\usepackage{url}
\usepackage{graphicx}

\usepackage{graphics}

\graphicspath{{./Figure}}
\usepackage{color}

\usepackage{amsmath,amssymb,amsfonts}%
\usepackage{amsthm}%
\usepackage{xcolor}%

\usepackage{orcidlink}

\usepackage{comment}
\usepackage{authblk}

\title{The Amerigo Vespucci as a traveling laboratory for studying the cosmic-ray fluxes at sea level}

\author[1,2]{Davide~Cerasole\,\orcidlink{0000-0003-2033-756X}\,}

\author[1]{Federica~Cuna\,\orcidlink{0000-0003-3414-9092}\,}

\author[1,2]{Gaia~De~Palma\,\orcidlink{0009-0002-5110-1445}\,}

\author[1]{Riccardo~Di~Tria\,\orcidlink{0009-0007-1088-5307}\,}

\author[1]{Leonardo~Di~Venere\,\orcidlink{0000-0003-0703-824X}}

\author[1]{Fabio~Gargano\,\orcidlink{0000-0002-5055-6395}}

\author[1,2]{Mario~Giliberti\,\orcidlink{0009-0007-2835-2963}\,}

\author[1]{Francesco~Licciulli\,\orcidlink{0000-0002-6955-0321}}

\author[1,2]{Antonio~Liguori\,\orcidlink{0009-0001-4240-6362}\,}

\author[1,3]{Pierpaolo~Loizzo\,\orcidlink{0000-0002-2404-760X}\,}

\author[1,2]{Francesco~Loparco\,\orcidlink{0000-0002-1173-5673}}

\author[1]{Leonarda~Lorusso\,\orcidlink{0000-0002-2549-4401}\,}

\author[1,*]{Mario~Nicola~Mazziotta\,\orcidlink{0000-0001-9325-4672}}

\author[1]{Giuliana~Panzarini\,\orcidlink{0000-0002-2586-1021}\,}

\author[1]{Roberta~Pillera\,\orcidlink{0000-0003-3808-963X}\,}

\author[1]{Davide~Serini\,\orcidlink{0000-0002-9754-6530}}

\affil[1]{Istituto Nazionale di Fisica Nucleare (INFN), Sezione di Bari, via Orabona, 4, Bari, 70125, Italy}

\affil[2]{Dipartimento di Fisica dell'Universit\`a  e del Politecnico di Bari, via Amendola 173, Bari, I-70126, Italy}

\affil[3]{Dipartimento di Fisica dell'Universit\`a di Trento,  via Sommarive, 14, Povo  (TN), I-38123, Italy}

\affil[*]{Corresponding Author, email: Marionicola.Mazziotta@ba.infn.it}

\date{}
\begin{document}

\maketitle

\thanks{Published in Scientific Reports (2025),
DOI: \url{https://doi.org/10.1038/s41598-025-24181-7}}

\begin{abstract}
We have installed and operated a plastic scintillator detector counter to measure the flux of cosmic radiation during the 2023$-$2025 tour of the historical vessel Amerigo Vespucci. The Vespucci is the oldest ship of the Italian Navy and serves as a training vessel for Navy cadets. During its tour, some experiments were hosted onboard the vessel, providing unique opportunities for scientists working in different fields. We installed our detector upon the Vespucci's departure from Darwin in early October 2024. The detector collected cosmic-ray data during the journey from Darwin to Trieste, where the worldwide tour ended in March 2025. After about one month of stop in Trieste, the ship continued its tour in the Mediterranean sea, and arrived in Genova on June 10, 2025. We performed measurements of the cosmic radiation reaching the sea level across a wide latitude range, from 15$^\circ$ S to about 45$^\circ$ N. The lowest rate (averaged over all azimuth angles) was measured at a geographic latitude of about 7$^\circ$ N, and was about 16\% less than the highest value, which was measured at Trieste, the northernmost location of the journey. Latitude effects on the cosmic radiation flux at sea level are due to the quasi-dipole geomagnetic field configuration, tilted by an angle of about 11$^\circ$ with respect to Earth's rotational axis.
\end{abstract}



\flushbottom

\maketitle

\thispagestyle{empty}

\section*{Introduction}
\label{sec1}
Cosmic rays (CRs) are particles that travel through space at nearly the speed of light. They are mostly protons ($\approx$ 90\%), but also include heavier atomic nuclei ($\approx$ 9\%), electrons/positrons, photons and neutrinos ($\approx$ 1\%)~\cite{ParticleDataGroup:2024cfk}.
The Earth's atmosphere prevents most of the radiation from the outer space from reaching the ground. 
When CRs enter the atmosphere, they collide with atmospheric molecules, producing cascades of secondary particles, including electrons/positrons, muons and neutrinos, that can reach the Earth's surface.

Discovered in 1912 by Victor Hess (measurements in the atmosphere)~\cite{Hess:1912srp} and by Domenico Pacini (underwater measurements)~\cite{Pacini:1912jqn}, CRs have been extensively studied for over a century.
The study of CRs led to major discoveries in particle physics, including the existence of positrons (antimatter)~\cite{Anderson:1933mb} and muons, and contributed to our understanding of subatomic particles. V. Hess was awarded the Nobel Prize in 1936 together with C. D. Anderson.

Soon after their discovery, scientists believed that CRs entering the atmosphere were high-energy gamma rays (neutral particles). In this case the Earth's magnetic field would not have any effect on their intensity. Few years before, C. St{\"o}rmer~\cite{stormer1907,stormer:jpa-00242218} developed a theory describing the motion of energetic charged particles in the Earth's dipole magnetic field to explain the phenomenon of aurorae borealis. However, the geomagnetic effects on CRs were discovered accidentally in the 1930s  by J. Clay~\cite{clay1927,clay1928}, while traveling on a ship from Java to the Netherlands. Clay found that ionization in the atmosphere increased with latitude, and this effect shows that at least a part of CRs must consist of charged particles.

The problem of measuring the geomagnetic effects on CRs and their correct interpretation was recognized after a public discussion between two Nobel Prize winners, Robert Millikan, who supported the gamma-ray hypothesis, and Arthur Compton, who supported the charged particle hypothesis. In 1932 Compton organized eight expeditions for measuring the CR intensity at several points at different latitudes and altitudes~\cite{Compton1933}. From the data collected in this campaign, the latitude effect on the CR rate on the Earth surface was recognized. The CR rate increased with increasing latitude, with a flattening above (below) $40^{\circ} - 50^{\circ}$ N (S).

The geomagnetic field can be approximately described as a dipole tilted by an angle of about 11$^\circ$ with respect to Earth's rotational axis, with the magnetic South pole close to the Earth's geographic North pole and the magnetic North pole close to the Earth's geographic South pole. Charged particles and nuclei approaching the Earth are deflected by the geomagnetic field, and the effect becomes relevant at low energies (or momenta). 
In particular, St{\"o}rmer found that, for some momenta and directions, CRs entering in the atmosphere are deflected or trapped, preventing them from reaching specific regions of the Earth’s surface. These forbidden regions are areas where CRs are blocked due to the Earth's magnetic field strength and orientation. The minimum momentum needed for a CR particle to reach a given location at the Earth is referred to as ``geomagnetic cutoff''. Usually, the geomagnetic cutoff is expressed in terms of rigidity (instead of momentum), is indicated by $\mathcal{R}_c$ and is expressed in units of GV, meaning that a given particle with a charge $Z$ (in units of the elementary charge $e$) must have a momentum $p$ (in units of GeV/c) not below $Z \mathcal{R}_c$ (for protons or electrons the rigidity value is equivalent to the momentum value, since Z=1). 

Before reaching the Earth, CRs must travel within the solar system, where their motion is influenced by the solar magnetic field. The CR intensity in the solar system is therefore affected by the variations of the solar magnetic field, which depends on the solar activity. The solar magnetic field has a complex and dynamic structure and is produced by the moving plasma (ions and electrons) inside the Sun. Places where very intense magnetic lines of force break through the Sun's surface are known as sunspots, and their number also depends on the solar activity, following a 11-year cycle. The sunspot number is commonly used as an indicator of the solar activity and is continuously monitored. On October 2024 the Sun reached the maximum activity of Solar Cycle 25, and the polarity of the solar magnetic field was reversed. 

The solar activity is characterized by transient events, referred to as ``solar flares'', which consist of intense and short duration emissions of electromagnetic radiation, usually coming from the most active regions of the Sun's surface, that may affect the CR intensity at Earth.
A ``Forbush decrease''~\cite{PhysRev.51.1108.3} is a rapid decrease in the observed cosmic-ray intensity following a Sun's coronal mass ejection (CME). This happens because the magnetic field carried by the solar wind plasma pushes some CRs away from the Earth. Auroras are the result of disturbances in the Earth's magnetosphere caused by the solar wind. Major disturbances result from enhancements in the speed of the solar wind from coronal holes and coronal mass ejections.

The Amerigo Vespucci is a tall ship of the Italian Navy (Marina Militare) named after the explorer Amerigo Vespucci. It is currently used as a training ship and it is well known as the most beautiful ship in the world. The Vespucci frequently carries out training cruises in the Mediterranean Sea, but it also sailed around the world to North and South America, and navigated across the Pacific Ocean.

On its last 2023$-$2025 world global tour, the Vespucci route covered a wide range of latitudes, in particular around the equator region, where the geomagnetic cutoffs are higher. A detector to measure the rate of secondary CRs at sea level
was installed on board of the Amerigo Vespucci vessel on Oct 4, 2024 at Darwin (Australia) and collected data until Mar 4, 2025, when the Vespucci ended its tour in Trieste (Italy). The detector consisted of two overlapping plastic scintillator tiles read out with silicon photomultipliers (SiPMs). The signals in the detector are due to the charged particles produced in air shower cascades initiated by CR particles in the upper layers of the atmosphere (20-30 km of altitude) which can reach the sea level (mainly positive/negative muons and a reduced fraction of electrons/positrons).

We studied the dependence of the CR flux on the longitude and latitude, due to changes in the Earth's magnetic field.
Our data are also sensitive to the solar activity, and we were able to measure the CR flux variations due to solar flares. To increase the accuracy of our measurements we equipped our detector with sensors to monitor the temperature, pressure, humidity, position, inclination and local magnetic field. 

After the arrival of the Vespucci in Trieste, on March 6, 2025 we moved our detector in a different place of the ship. With the detector in its new location, we then measured the CR flux in the final part of the Vespucci Mediterranean tour from Trieste (Mar 7, 2025) to Genova (Jun 11, 2025). The results of this second campaign are also reported. 

We have also used the data of the Oulu neutron monitor (NM) as a reference for both data taking campaigns on the Vespucci ship. NMs are detectors placed at different locations on the Earth~\cite{nmstation} that provide accurate measurements of the intensity of CRs striking the upper atmosphere and its variations with time~\cite{2010ASPC..424...75M}. A NM detector measures the neutron component of secondary CRs. The count rate observed by a NM is proportional to the flux of primary CRs impinging on the upper atmosphere. The network of NMs on the Earth surface allows a real-time monitoring of the CR fluxes around the planet. In particular, NMs located at high latitudes allow the study of the lowest energy component of CRs, since the geomagnetic cutoff near the poles is very low. The Oulu station, operated by the Sodankyla Geophysical Observatory of the University of Oulu (Finland), is located at geographic latitude 65.05$^\circ$ N and geographic longitude 25.47$^\circ$ E and is one of the northernmost detectors in the NM network.

In addition, from March 2025, we have also used a second CR muon counter installed at the INFN Bari~\cite{Pillera:2023uyr} and similar to the one installed on the Vespucci.

\section*{Results}
\label{sec2}

\subsection*{World tour}

The cosmic-ray rate at sea level was measured onboard the Vespucci vessel during its journey from Darwin (Oct 4, 2024) to Trieste (Mar 4, 2025) in the geographic (longitude, latitude) interval ranging from (130.85$^\circ$ E, 12.47$^\circ$ S) to (13.77$^\circ$ E, 45.63$^\circ$ N).
The detector does not have tracking capability and provides only information about the passage of particles in the two tiles. The measured rate is obtained by dividing the number of events collected in each data acquisition run for its duration. The rate has been finally corrected taking into account the detector inclination and the barometric effects (details are given in the ``Methods'' section).

\begin{figure}[!htb]
\centering
\includegraphics[scale=0.4]{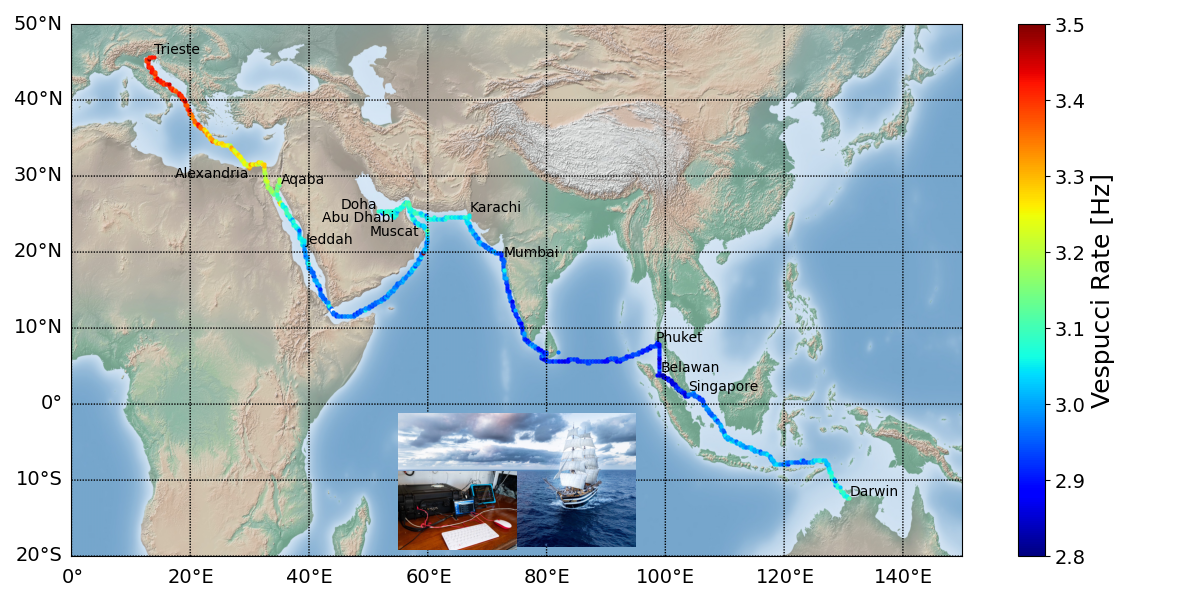}
\caption{Cosmic-ray rate map measured on board the Vespucci vessel during its journey from Darwin to Trieste. The color gradient shows the rate values from 2.8 Hz (dark blue) to 3.5 Hz (dark red) with  grid spacing resolution of $0.25 \times 0.25 ~deg^2$. The cities on the map refer to the main stops of the Vespucci. The map was generated using the Matplotlib Basemap toolkit~\cite{Hunter:2007} version 1.4.0 (\url{https://matplotlib.org/basemap/1.4.0/users/index.html}).}
\label{fig:vesmap}
\end{figure}

\begin{figure}[!htb]
\centering
\includegraphics[scale=0.3]{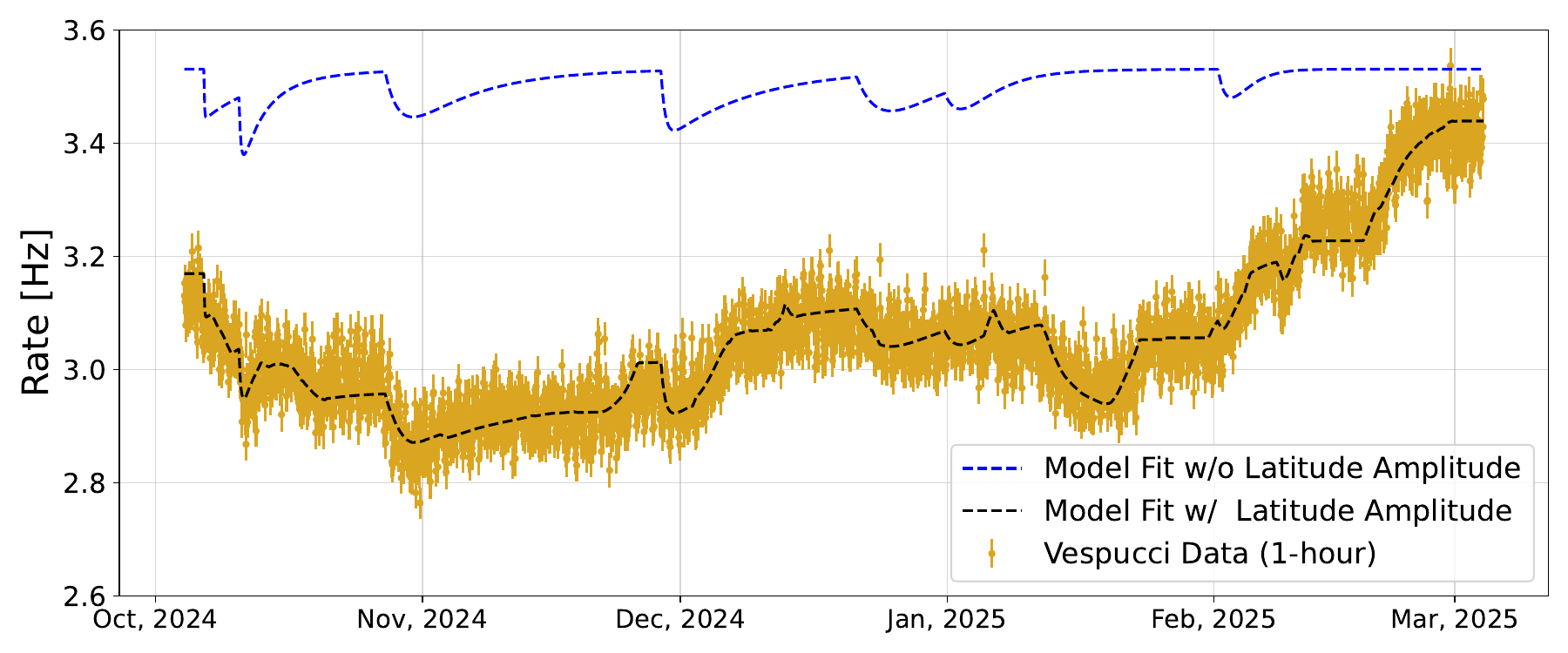}
\includegraphics[scale=0.3]{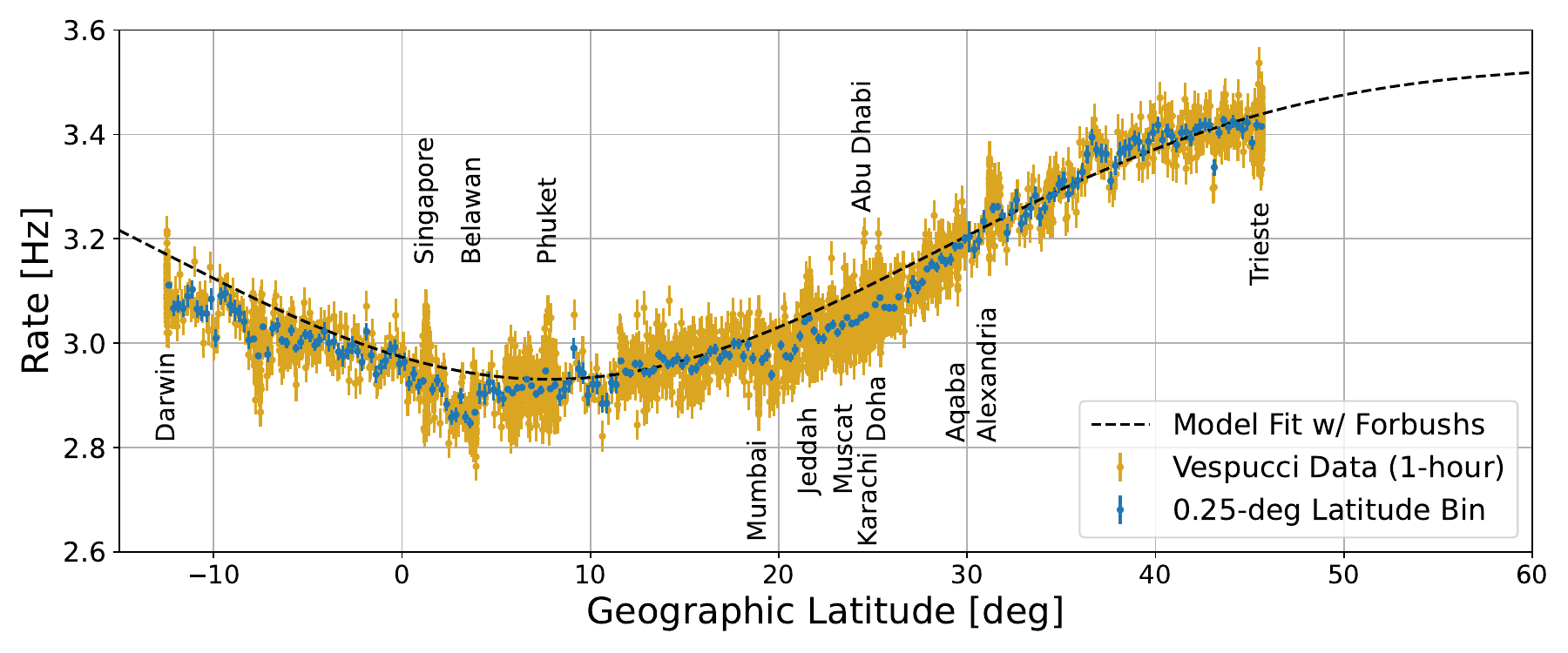}
\includegraphics[scale=0.3]{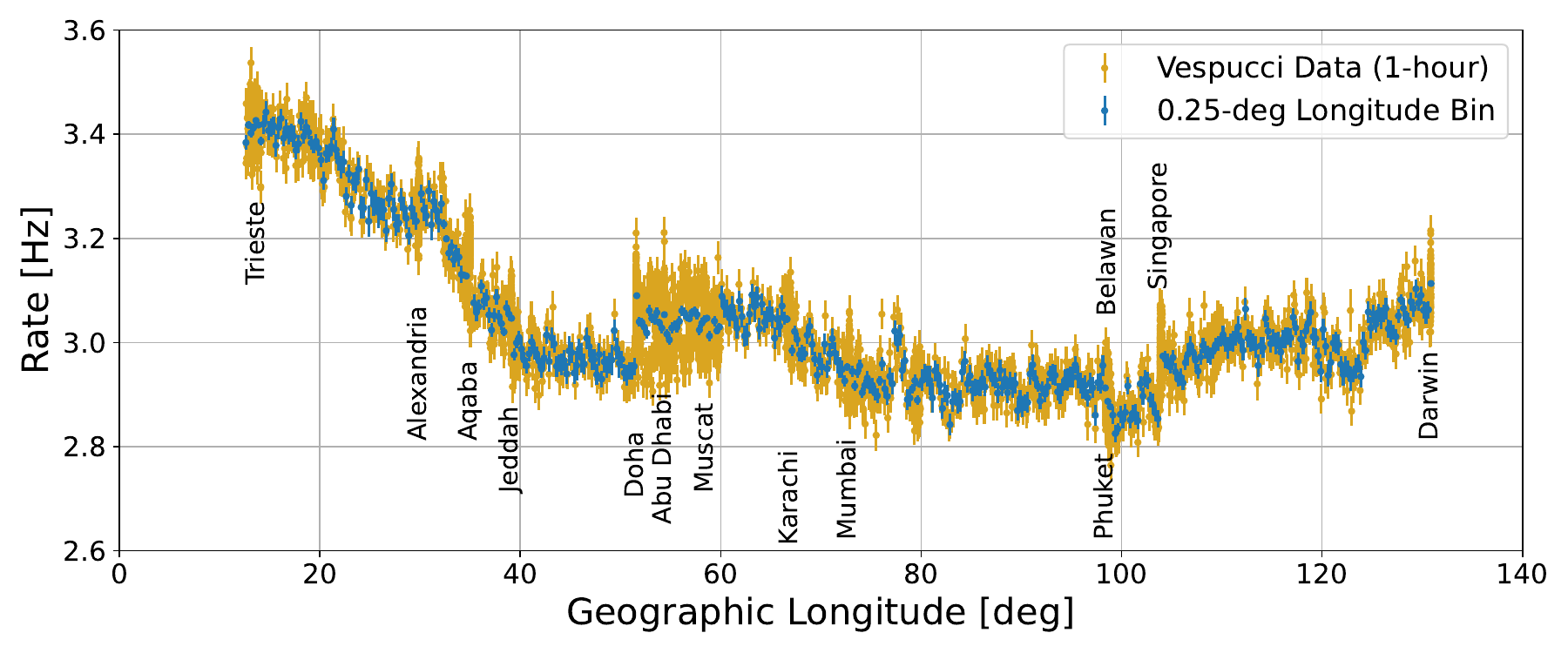}
\caption{Top panel: Cosmic-ray rates measured by the detector on board the Vespucci as a function of time since October 4, 2024. Each golden marker corresponds to a 1-hour time interval (UTC time). The dashed black line shows the results of the best fit with Eq.~\ref{eq:rfit}. The blue dashed line shows the result of the best fit neglecting the latitude variations (Eq.~\ref{eq:rfit} with the parameter $\alpha=0$). Middle panel: Cosmic-ray rates as a function of the geographic latitude. The dashed black line is the best fit results with the latitude term of Eq.~\ref{eq:rfit}. Bottom panel: Cosmic-ray rates as a function of the geographic longitude. Each golden marker corresponds to a 1-hour time interval. The blue points indicate the average rates evaluated in $0.25^{\circ}$ latitude (longitude) bins. The vertical bars represent the statistical errors.} 
\label{fig:vestimlat}
\end{figure}

The rates of secondary CRs measured by the detector as a function of the geographic position, time and geographic latitude (longitude) during the whole journey are shown in Fig.~\ref{fig:vesmap} and Fig.~\ref{fig:vestimlat}, respectively. 
Rate variations, due to both latitude and solar activity effects, are clearly visible. 
In the top panel of Fig.~\ref{fig:vestimlat}, long term variations are mainly due to the latitude change, while rapid decreases are associated to solar activity. In particular, we found a minimum in the rate in the area around Singapore, Belawan and Phuket, at a latitude of about 7$^\circ$ N. In this region the geomagnetic field configuration yields the maximum rigidity cutoff and consequently the CR flux is minimum. In the middle of January 2025, we observed a well defined minimum in the rate not associated to any significant solar activities: this occurred during the circumnavigation of the Arabian Peninsula, from Muscat to Jeddah, and is a clear evidence of the latitude effect on the CR fluxes at sea level. 

The lowest rate measured near the geographic latitude of about 7$^\circ$ N (averaged over all azimuths) was about $16\%$ less than the value measured at Trieste. The measured CR rate seems to saturate just above 40$^\circ$ latitude (or below 5 GV in vertical cutoff rigidity). However, Trieste is the northernmost position of the Vespucci 2023$-$2025 world tour, and therefore we did not have the opportunity to take measurements above this latitude.

The CR rates measured on the Vespucci vessel are affected by possible variations due to the solar activity and latitude effects. Indeed, from Oct 2024 the Sun reached its maximum activity for the Cycle 25, characterized by several intense solar flares and coronal mass ejections (see for instance ~\cite{cmeoct24}). The rate data as a function of time have been fitted with the following function:

\begin{equation}
\tilde{R}(t) = \tilde{R}_0 \underbrace{\left(1 - \sum^N_{j=1} \tilde{F}_j(t) \right)}_\text{Forbush term} \times \underbrace{ \tilde{L}(\lambda) }_\text{Latitude term}
\label{eq:rfit}
\end{equation}
Here $\tilde{R}_0$ is the rate that would be observed at polar latitudes without solar activity effects. The Forbush term accounts for all the $N$ Forbush decreases observed during the journey. The effect of the $j$-th Forbush decrease on the rate is modeled with the function:

\begin{equation}
\tilde{F}_j(t) =  a_j \left(e^{-(t-t_j)/b_j} - e^{-(t-t_j)/c_j}\right)  H(t-t_j)
\label{eq:eq2}
\end{equation}
where $t_j$ is the time when the $j$-th Forbush decrease started, $H(t-t_j)$ is the Heaviside step function, $a_j$ is the amplitude decrease, $b_j$ and $c_j$ are the rise and fall time, respectively.
Finally, the latitude term

\begin{equation}
\tilde{L}(\lambda)  = 1 - \alpha \, cos^{\gamma} (\lambda - \lambda_0) 
\label{eq:eq13}
\end{equation}
accounts for the latitude effect. Here $\lambda$ is the latitude position at the current time $t$, $\alpha$ is the amplitude of the latitude variations, $\lambda_0$ is the latitude corresponding to the minimum rate and $\gamma$ accounts for the plateau width in latitude.


\begin{figure}[!t]
\centering
\includegraphics[width=\textwidth]{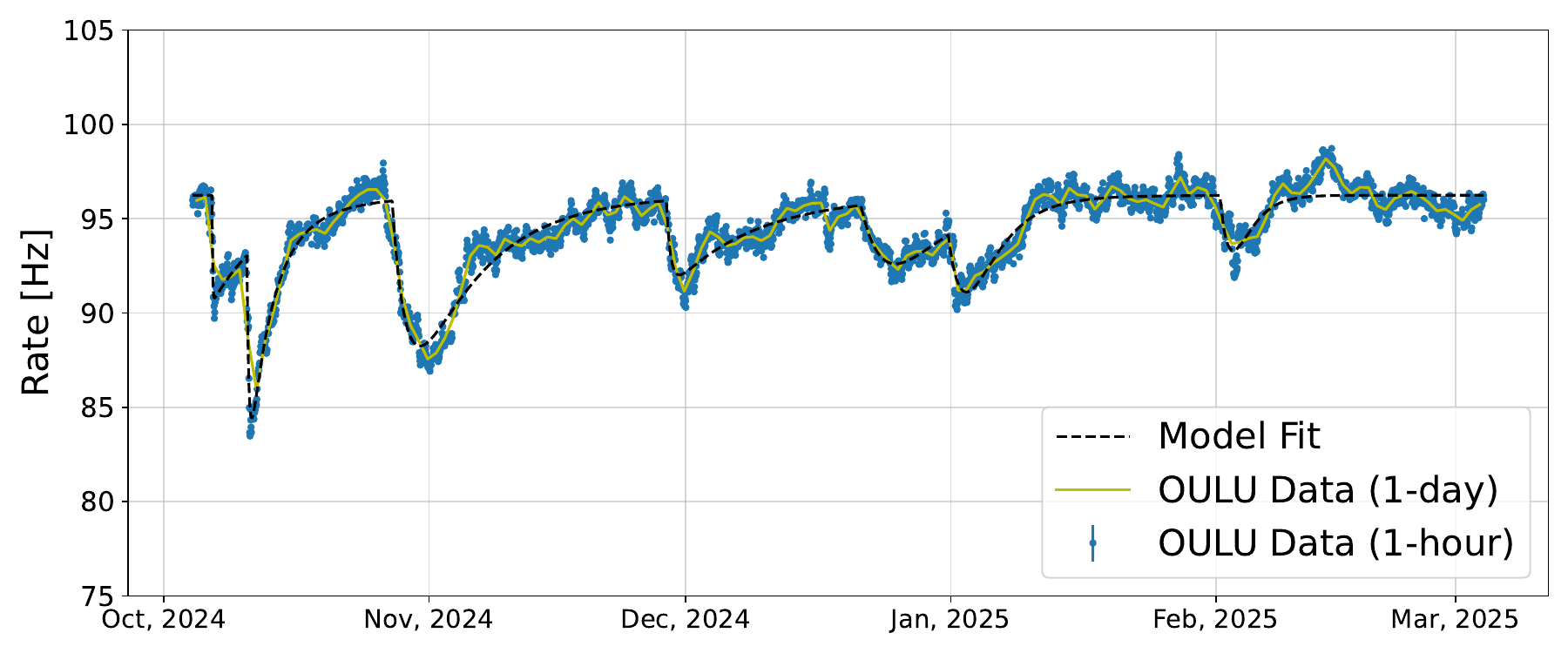}
\caption{Cosmic-ray rates measured by the Oulu detector as a function of time since October 4, 2024 (UTC time). The blue points show the data in 1-hour time intervals. The dashed black line indicates the results of the best fit with Eq.~\ref{eq:rfit}, without including the latitude effect. The yellow line shows the rates in 1-day time intervals.}
\label{fig:oulu}
\end{figure}

\begin{figure}[!t]
\centering
\includegraphics[scale=0.5]{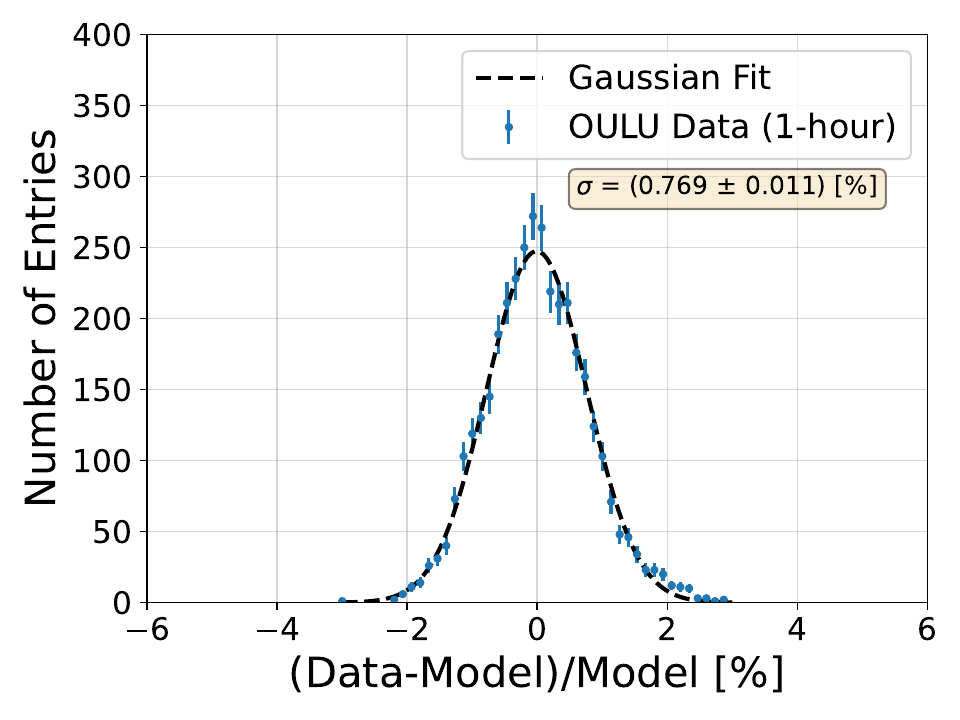}
\includegraphics[scale=0.5]{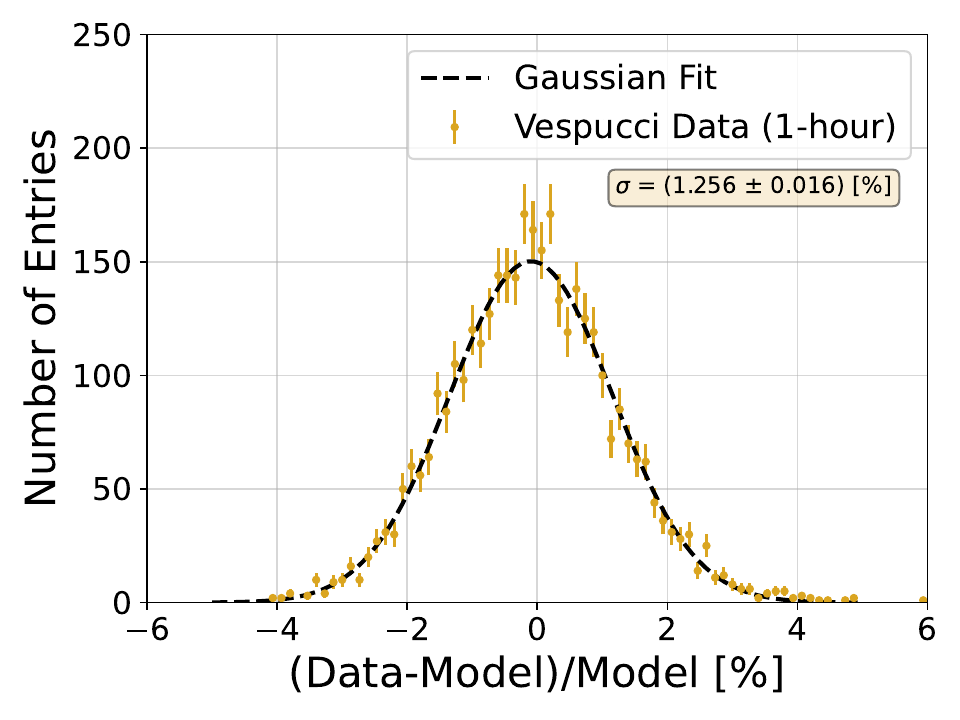}
\caption{Distributions of fractional residual (in units of \%) between the observed rate and the best fit model (colored points) with the Oulu detector (left panel) and with the Vespucci detector (right panel). The dashed black lines show Gaussian fits with the best-fit sigma value in the text box superimposed. The vertical bars indicate statistical uncertainties.}
\label{fig:res}
\end{figure}

Setting $\alpha=0$ in the latitude term, Eq.~\ref{eq:rfit} yields a constant rate function modulated by possible Forbush decreases. On the other hand, setting $a_{j}=0$, Eq.~\ref{eq:rfit} yields only the latitude variations of the rate. We remark here that the latitude term is just a phenomenological model that would describe adequately the observed effect, and is not based on any physical consideration.

To identify possible Forbush decreases and the corresponding start times, we have first studied the rates measured by the Oulu NM. 
The Oulu data have been fitted with Eq.~\ref{eq:rfit}, neglecting the latitude dependence, i.e. setting $\tilde{L}(\lambda)=1$ ($\alpha=0$). The results are summarized in Fig.~\ref{fig:oulu}, which shows the CR rate measured by the Oulu NM as a function of time with the fit function superimposed. The rate exhibits sudden variations due to the solar activity, corresponding to Forbush decreases. Fluctuation on shorter time scales are also evident, probably due to night/day effect and solar activity. We identify seven  main Forbush events. 
The start times of these Forbush events, evaluated from the fits, are: 
Oct 6, 2024 at 14:53 UTC; 
Oct 10, 2024 at 15:40 UTC; 
Oct 27, 2024 at 15:46 UTC; 
Nov 28, 2024 at 18:02 UTC;
Dec 21, 2024 at 11:42 UTC;
Dec 31, 2024 at 16:58 UTC and
Feb 1, 2025 at 10:06 UTC. The 1-sigma uncertainties of the Forbush start-time range from about 5 min to about 27 min.

\begin{figure}[!th]
\centering
\includegraphics[scale=0.5]{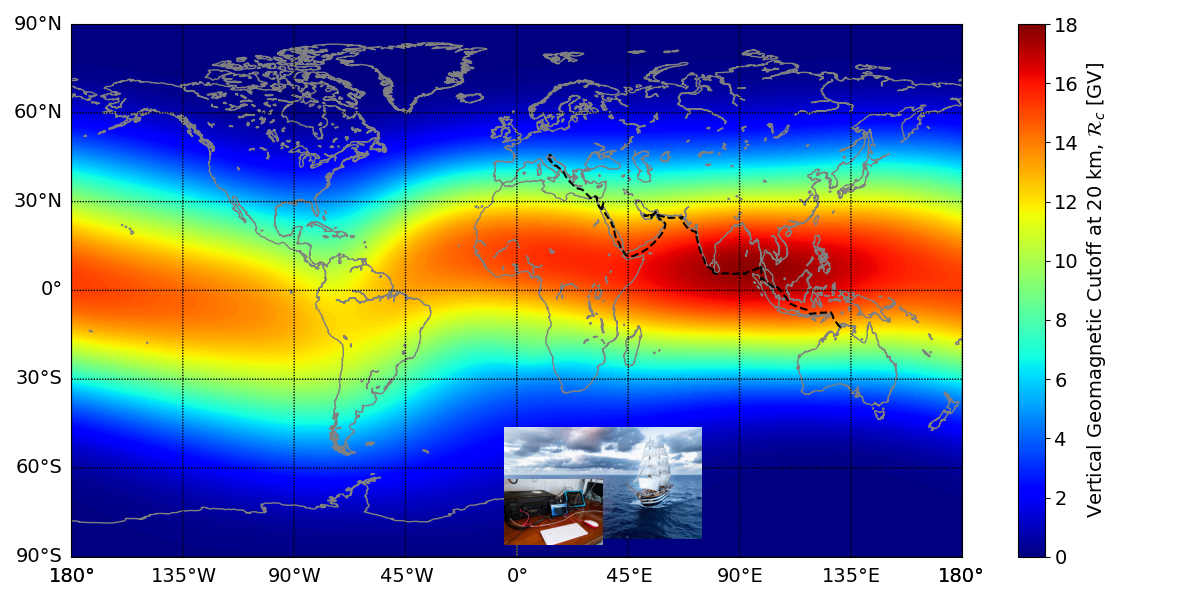}
\caption{Vertical cutoff rigidity evaluated from the IGRF geomagnetic field model at the epoch 2024.9 on a $1^\circ \times 1^\circ$ world grid. The dashed black line shows the route of the Vespucci vessel from Darwin to Trieste. The map was generated using the Matplotlib Basemap toolkit~\cite{Hunter:2007} version 1.4.0 (\url{https://matplotlib.org/basemap/1.4.0/users/index.html}).}
\label{fig:rcmap}
\end{figure}

We did not observe any significant time lag between the Oulu NM and the Vespucci  data for the Forbush events. In addition, the time dependence of the count rate (fall and rise time) is also very similar between the two data sets for all Forbush events. From the fit of the Oulu data we have therefore extracted the initial times $t_j$, the rise times $b_j$ and the fall times $c_j$ of the 7 Forbush decreases. We have then fitted the data observed on board the Vespucci including the possible variations of the rate with the latitude (Eq.~\ref{eq:rfit}), and fixing the parameters $t_j$, $b_j$ and $c_j$ for the seven Forbush events to the best-fit values of the Oulu data, leaving the amplitude parameters $a_j$ free. The best-fit parameter values are $\tilde{R}_0= (3.531 \pm 0.006)$ Hz, $\alpha=(0.170 \pm 0.001)$, $\lambda_0=(7.80 \pm 0.09)^\circ$ N and $\gamma=(7.95 \pm 0.16)$, respectively.

The results of the best-fit procedure on the Vespucci data are shown with dashed lines in Fig.~\ref{fig:vestimlat}. In particular we also show the observed rate as function of time, disentangling the latitude variation effects (blue dashed line on the top panel of Fig.~\ref{fig:vestimlat}).
The rate decreases for the seven Forbush events  measured by the Oulu NM are in a range from 
$\approx 3.1\%$ to $\approx 12.3\%$, while the corresponding decreases measured by the Vespucci CR detector are in a range from $\approx 1.1\%$ to $\approx 4\%$. Typically, the rate drops in the Vespucci muon detector are a factor $\approx 3$ lower than those in the Oulu NM detector. We also remark that the second Forbush decrease occurred 4 days after the first one, and therefore there is an overlap between their effects.

Fig.~\ref{fig:res} shows, for the Oulu NM data (left panel) and for the Vespucci data (right panel) the distributions of the fractional residuals (in percentage units) between the measured rates and the predictions of the best-fit model. In both cases the residual distribution is well fitted by a gaussian with null mean, with $\sigma \approx 0.76\%$ for the Oulu NM data and $\sigma \approx 1.2\%$ for the Vespucci data.

\begin{figure}[!t]
\centering
\includegraphics[scale=0.7]{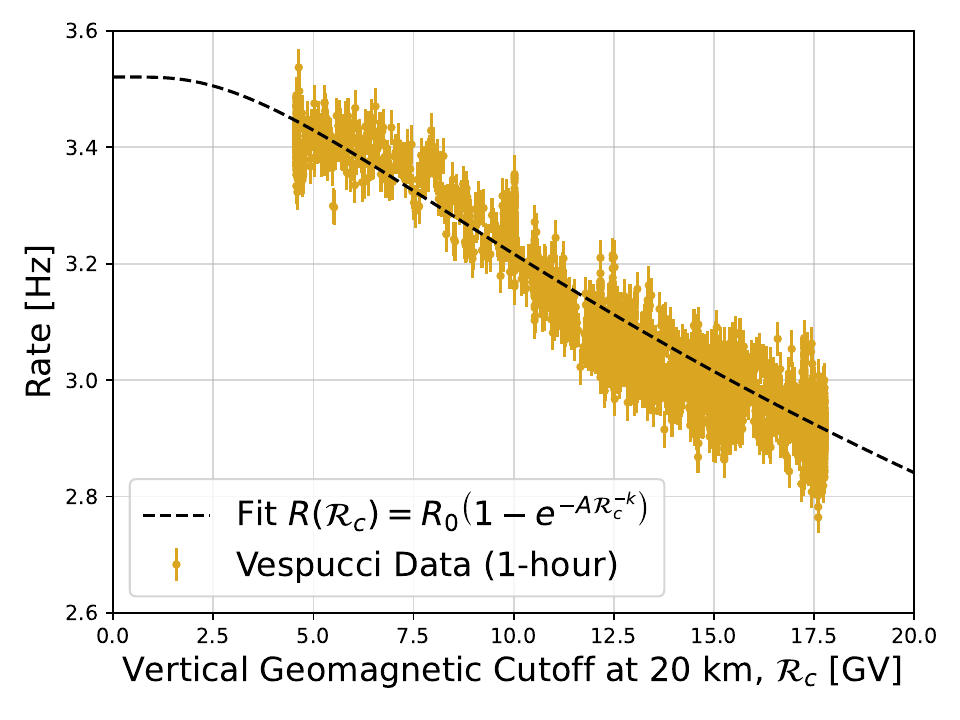}
\caption{Cosmic-ray rate measured by the Vespucci detector as a function of the vertical geomagnetic cutoff. Golden data points correspond to 1-hour time intervals. The vertical bar indicate statistic errors. The dashed black line shows the results of the best fit with Eq.~\ref{eq:rfitrc}.}
\label{fig:vesrc}
\end{figure}

We have then evaluated the vertical geomagnetic cutoff along the route traveled by the Vespucci from Darwin to Trieste using the International Geomagnetic Reference Field (IGRF)~\cite{alken2021international,Laundal2024-fl} at 20 km altitude, where CRs impinging on the Earth interact with the atmospheric nuclei. The rigidity cutoff depends on the arrival direction at a given Earth's location, in the following we will refer to the vertical cutoff evaluated for particles coming from the zenith directions. The map of the geomagnetic cutoff as a function of the geographic coordinates is shown in Fig.~\ref{fig:rcmap}. The secondary CR rate measured at sea level onboard of the Vespucci vessel as a function of the vertical geomagnetic cutoff $\mathcal{R}_c$ is shown in  Fig.~\ref{fig:vesrc}. The observed data rate on the Vespucci can also be fitted with the function: 

\begin{equation}
\tilde{R}(t) = \tilde{R}_0 \underbrace{\left(1 - \sum^N_{j=1} \tilde{F}_j(t) \right)}_\text{Forbush term} \times \underbrace{ \tilde{R}(\mathcal{R}_c) }_\text{Cutoff term}
\label{eq:rfitrc}
\end{equation}
where $\mathcal{R}_c$ is the vertical cutoff rigidity (in units of GV) as a function of time along the Vespucci route.
The function $\tilde{R}(\mathcal{R}_c)$ is chosen following Ref.~\cite{dorman2009cosmic} (``Dorman function'') and is given by: 

\begin{equation}
    \tilde{R}(\mathcal{R}_c) = \left(1 - e^{-A \mathcal{R}_c^{-k}}\right) 
    \label{eq:rc}
\end{equation}
where $A$ and $k$ are two fit parameters introduced to describe the saturation of the rate with decreasing geomagnetic cutoff.  
The best-fit parameters are $A=(9.21 \pm 0.22)$ and $k=(0.575 \pm 0.007)$, respectively. The best fit function is shown in Figure~\ref{fig:vesrc} as a dashed black line superimposed to the experimental data. 

\subsection*{Mediterranean Tour}
\label{secA:MedTour}

After the 2023-2025 world tour terminated in Trieste, the Vespucci continued its journey in the Mediterranean sea making stops in many Italian cities, in Albania and in Malta, up to Genova, where it concluded its journey on June 10, 2025. In Trieste the detector counter was moved in a different room, very close to the initial one where it was installed during the journey from Darwin to Trieste, and it continued the data taking until Genova. We found that the rate measured by the detector in the new position was about $3\%$ lower than the rate measured in the previous position, probably due to additional materials above the counter. 

\begin{figure}[!th]
\centering
\includegraphics[scale=0.45]{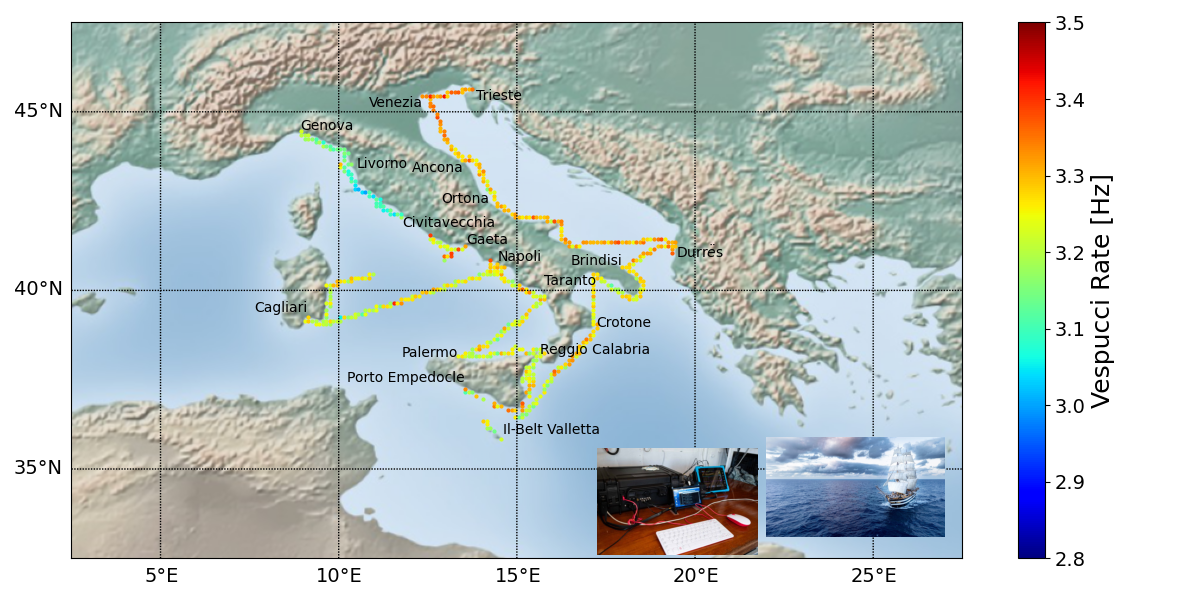}
\caption{Cosmic-ray rate map measured on board the Vespucci vessel during its journey from Trieste to Genova. The color gradient shows the rate values from 2.8 Hz (dark blue) to 3.5 Hz (dark red) with  grid spacing resolution of $0.1 \times 0.1 ~deg^2$. The cities on the map refer to the main stops of the Vespucci. The map was generated using the Matplotlib Basemap toolkit~\cite{Hunter:2007} version 1.4.0 (\url{https://matplotlib.org/basemap/1.4.0/users/index.html}).}
\label{fig:vesmapTR}
\end{figure}

\begin{figure}[!t]
\centering
\includegraphics[width=1\textwidth]{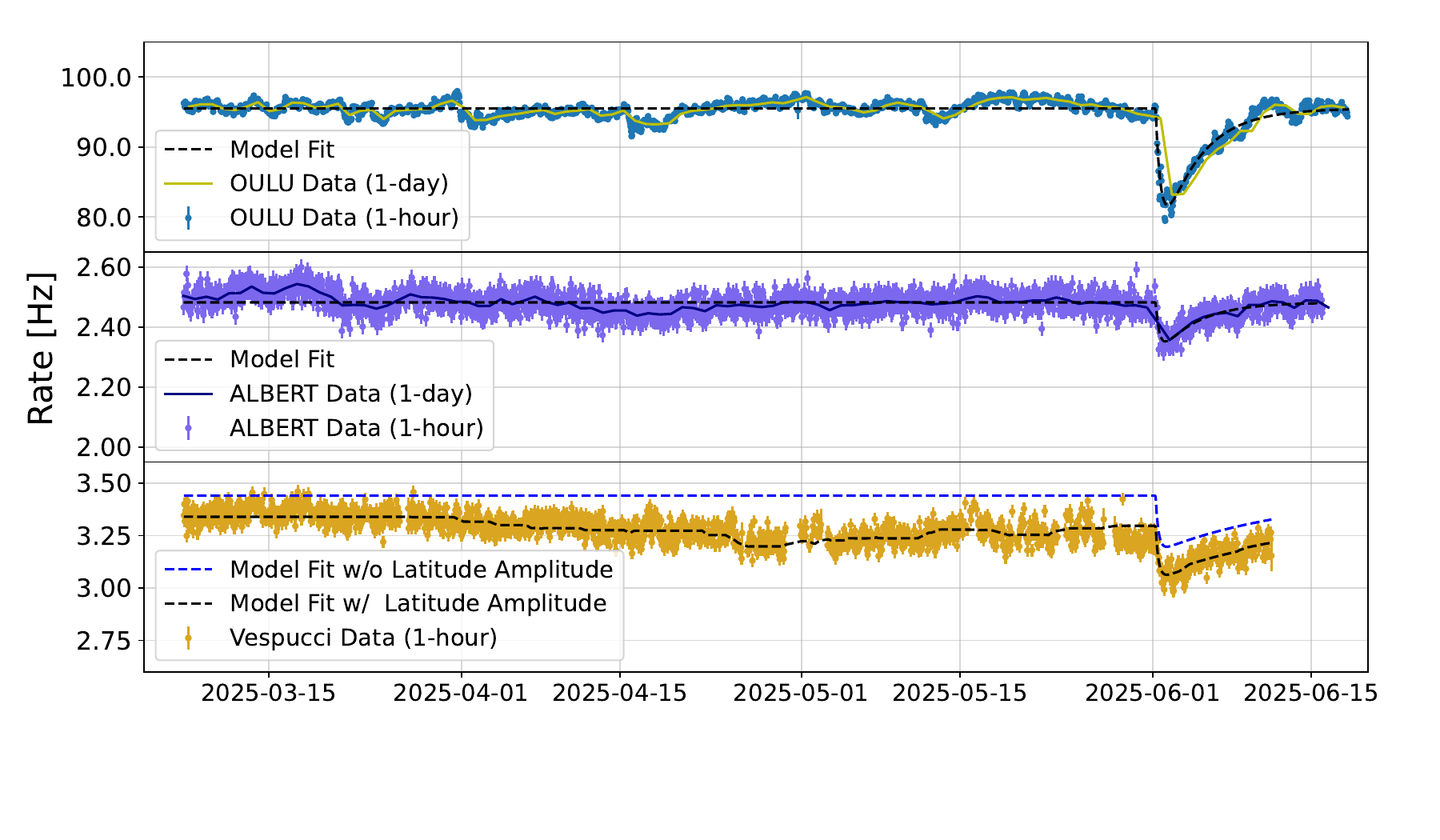}
\includegraphics[width=0.95\textwidth]{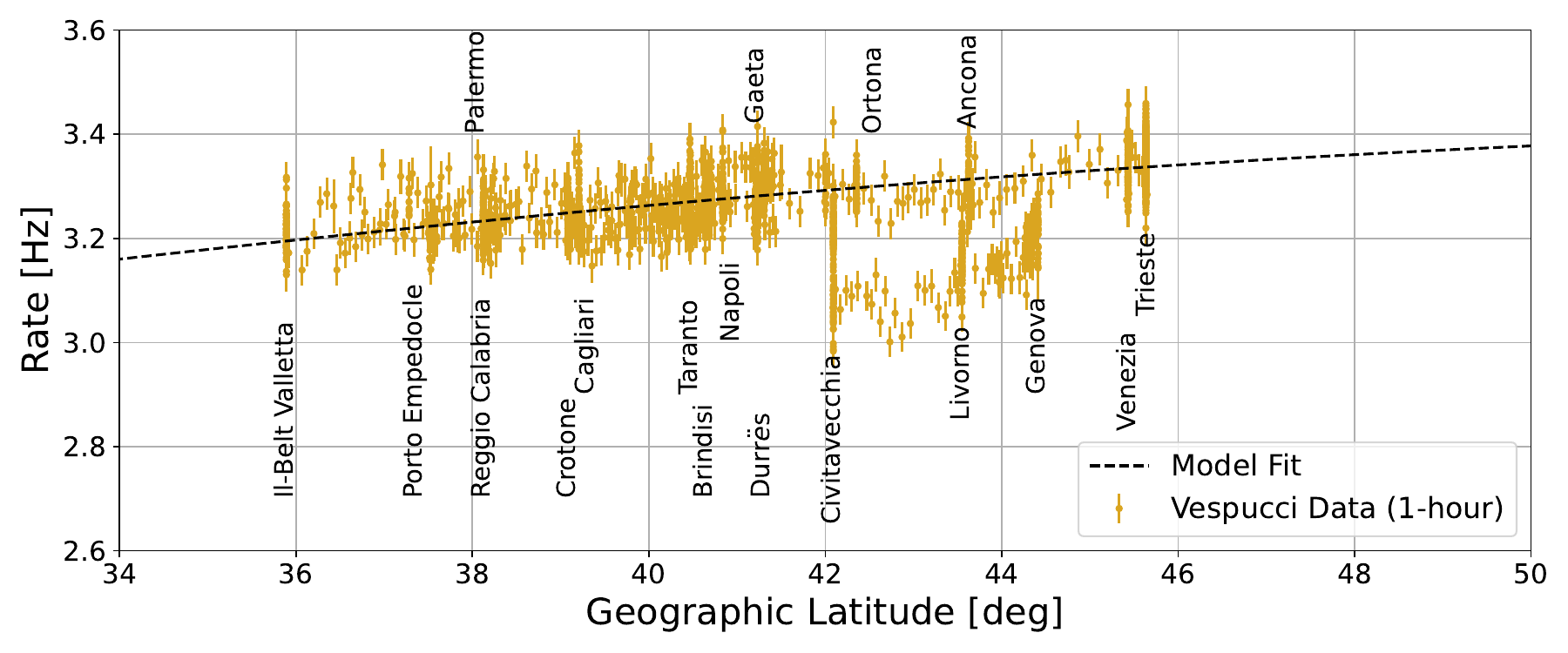}
\caption{Top panels: CR rates measured by the Oulu NM, by the ``ALBERT'' detector at INFN Bari and by the CR detector onboard the Vespucci ship as a function of time from March 7, 2025 to June 11, 2025. Bottom panel: CR rate as a function of the geographic latitude measured on board of the Vespucci ship. The vertical bars represent the statistical errors.}
\label{fig:vestimlatTR}
\end{figure}

\begin{figure}[!th]
\centering
\includegraphics[scale=0.7]{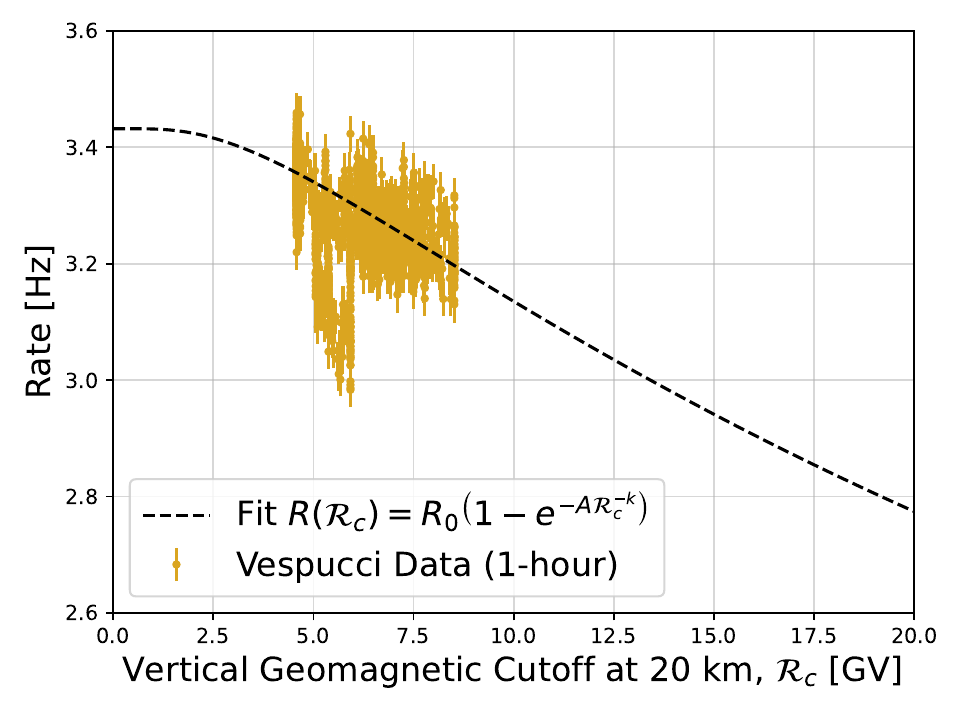}
\caption{Cosmic-ray rate measured by the Vespucci detector from Trieste to Genova as a function of the vertical geomagnetic cutoff. Golden data points correspond to 1-hour time intervals. The points with lower rate in the cutoff rigidity range  $\mathcal{R}_c$ between 5 and 6 GeV correspond to the Forbush event at the end of May, 2025. The vertical bar indicate statistical errors. The dashed black line shows the results of the best fit with Eq.~\ref{eq:rfitrc}.}
\label{fig:vesrcTR}
\end{figure}

The rate of secondary CRs measured by the Vespucci detector as a function of the geographic position during the journey from Trieste to Genova is shown in Fig.~\ref{fig:vesmapTR}. The rate measured by the Oulu NM in the same time period is also shown in the top panel of Fig.~\ref{fig:vestimlatTR}. The Oulu data do not show any significant Forbush decreases until the end of May, when we find a significant decrease of the rate due to a solar flare event occurred between May 28 and 29. At that time, the Vespucci was in Civitavecchia, where we also observed a decrease of the count rate. 
The start times of this Forbush event, evaluated from the fit of the data with eq.~\ref{eq:rfit}, is June 1, 2025, 6:13 UTC; the fall and rise times were (0.4668 $\pm$ 0.006) days and  (3.338 $\pm$ 0.014) days, respectively, and the coefficient $a_j$ in Eq.~\ref{eq:eq2} describing the amplitude of the rate decrease was of about 0.23.

After including this Forbush event in the fit, the fluctuations around the average value are still within $1\%$. In Fig.~\ref{fig:vestimlatTR} we show the rate measured by the counter on the Vespucci as a function of time (last plot of the top panel) and as a function of latitude (bottom panel). The rate measured on the Vespucci also does not exhibit any particular feature, except for the Forbush event at the end of May and the latitude effect. Consequently, the Vespucci data have been fitted with Eq.~\ref{eq:rfit}, where the parameter $\tilde{R}_0$ is free, while the parameters $\alpha$, $\beta$ and $\lambda_0$ are fixed to the best value found with the data collected from Darwin to Trieste. The results of the fit are shown with the dashed black line in Fig.~\ref{fig:vestimlatTR}. The coefficient describing the amplitude decrease of the June Forbush event is of about 0.08 (i.e. about 1/3 of the corresponding decrease found in the Oulu NM data). The residuals with respect the best fit model still exhibit a gaussian-like distribution with null mean and $\sigma \approx 1.2\%$. 
In Fig.~\ref{fig:vestimlatTR} we also show the rate measured with the CR detector ``ALBERT''~\cite{Pillera:2023uyr} installed at the INFN Bari. In this case we find that the amplitude decrease coefficient associated to the Forbush event is of about 0.08, similar to the result achieved with the Vespucci data. Moreover, the falling and rising times evaluated with the fit procedure are in agreement with the Oulu and Vespucci data. We remark here that the measured CR rates by the detector onboard of the Vespucci and by ALBERT are different because of different detector sizes and configurations.

The rate from Trieste, the northernmost location, to La Valletta, the southernmost position during the Vespucci Mediterranean tour, decreases of about $5\%$, in agreement with the same latitude decreases observed during the 2023-2025 World tour in the route Darwin-Trieste shown in bottom panel of Fig.~\ref{fig:vestimlat}.

Finally, the rate measured onboard of the Vespucci vessel from Trieste to Genova as a function of the vertical geomagnetic cutoff $\mathcal{R}_c$ is shown in  Fig.~\ref{fig:vesrcTR}. The observed data rate are still fitted with the Dorman function in Eq.~\ref{eq:rc} fixing the parameters $A$ and $k$ to the values obtained from the data from Darwin to Trieste. The results of the best fit procedure is also shown with the dashed line on Fig.~\ref{fig:vesrcTR}. 

\section*{Discussion}
\label{sec12}

After the first surveys conducted by Compton and Millikan in the 1930s, since the 1950s many other expeditions were organized to study the rate of CRs as a function of geographic latitude~\cite{dorman2009cosmic}. The experimental data collected in those expeditions are unique, as the geomagnetic field changes with time and consequently the cutoff rigidities across the Earth also change. 

In the previous expeditions the CR rate was measured mainly with neutron detectors, that are sensitive to the secondary nucleons produced in the CR showers initiated at the top of the atmosphere. In a few of them, counters (e.g. ionization chamber) were used that are sensitive mainly to the charged particles (e.g. muons, electrons, positrons) produced in the atmospheric cascades initiated by CRs. The minimum rate (CR equator) was observed around 7$^\circ$ N and 10$^\circ$ W~\cite{dorman2009cosmic}, in agreement with the current results. 

The Japanese Antarctic Research Expedition (JARE) projects performed CR observation with NMs in the Antarctica until early 1960s~\cite{1961175} and measured both the nucleon and muon component in the latitude range from about 25$^\circ$ N to about 70$^\circ$ S. They found a minimum of the radiation at about 7$^\circ$ S in geomagnetic latitude and a plateau starting from 35$^\circ$ S also in geomagnetic latitude (the geomagnetic latitude, or magnetic latitude, is similar to geographic latitude, except that it is defined by the axis of the geomagnetic dipole). It slightly depends on time and can be extracted from the IGRF, with a higher decrease of the nucleon component with respect to the muon component~\cite{Kodama1958}. 

In the Swedish-USA latitude surveys during 1955-1959 in connection with the International Geophysical Year, from Scandinavia to South Africa and back via Australia and Suez Canal, the minimum CR rate was found at a geographical latitude of 6.9$^\circ$ N and longitude 14$^\circ$ W~\cite{Pomerantz1958}, in agreement with our results.

The muon rate decrease with latitude (or equivalently with geomagnetic rigidity cutoff) is less pronounced than the corresponding decrease observed with NMs. This effect was also observed by the authors of Ref.~\cite{doi:10.1139/p56-001} with the  Labrador Canadian Naval Icebreaker into the Arctic, in the geomagnetic latitude range from 18$^\circ$ N to 89$^\circ$ N, through the North West Passage, and circumnavigating the North American Continent. In this case the plateau was observed starting at about 40$^\circ$ N (see also~\cite{Abbrescia:2023vop}). The asymmetry of the plateau in South and North Earth hemispheres is due to the configuration of the geomagnetic field, with the maximum vertical rigidity cutoff equator above the geographical one as shown in Fig.~\ref{fig:rcmap}. 

Similar results were obtained in latitude survey campaigns from 1994 to 2007 with a 3BN64 neutron monitor operated on two U.S. Coast Guard icebreakers, the Polar
Sea and the Polar Star, which traversed the Pacific Ocean from Seattle (USA) to McMurdo (Antarctica) and back in a 6-month voyage~\cite{Nuntiyakul_2014}. In this survey, the CR rate was measured in a geomagnetic cutoff rigidity range up to 15 GV, and the rate decrease was about $40\%$ higher than that measured with a particle counter similar to the one used in the present work. The calculated coefficient of the Dorman function $A$ ranges from 8.6 to 11.1, while $k$ ranges from 0.862 to 0.926, depending on the year. The value of the  parameter $k$ obtained from the Vespucci data is lower than the one obtained with NM data. This difference shows that the nuclear component of cosmic rays is attenuated faster by the atmosphere than the muon component.

A similar effect is also observed during the Forbush decrease events, where we find that the muon rate decrease is lower that the corresponding neutron rate decrease. These results demonstrate that muons are sensitive to primary CRs of higher energies than neutrons.

\section*{Conclusion}
\label{sec:concl}

In the present work we measured the CR rate at sea level with a plastic scintillator telescope installed onboard the Vespucci vessel, finding similar results as in the past historical expeditions. We found the minimum rate at a latitude of about 7$^\circ$ N, with a decrease of about 16\% with respect to the rate measured at the northernmost position in Trieste. We also identified seven Forbush decreases in our data, due to the intense solar activity during the measurement period.

After the $2023 - 2025$ world tour until to Trieste, the Vespucci continued its journey in the Mediterranean sea, making stops in many Italian cities, in Albania and in Malta, up to Genova where it concluded its journey on June 10, 2025. The measurements obtained during the trip from Trieste to Genova are in agreement with the results obtained in the journey from Darwin to Trieste.

CR observations are very useful in understanding the Earth's magnetospheric structure. In addition, they are also useful to study the solar activity during the 11-year cycles and in presence of solar flares with coronal mass ejection events. The long-term measurements performed onboard the Vespucci vessel, with a single instrument operating in stable conditions, allowed to obtain a precise data set with reduced systematic effects, covering a wide range of geographic (and geomagnetic) coordinates. Our findings are in agreement with previous observations with other instruments. In addition, during its journey, our detector has allowed an accurate monitoring of the solar activity, which was particularly enhanced, since the current solar cycle reached its maximum in this period. 

\section*{Methods}
\label{sec:met}

\subsection*{Detector description}
\label{secA:det}

The detector onboard the Vespucci consists of two overlapping 1 cm thick tiles of plastic scintillator. Each tile, with a cross section of $15\times15$ cm$^2$, is wrapped with a 25 $\mu m$ thick aluminized polyethylene foil, enclosed within two 50 $\mu m$ thick layers of light-tight black DUPONT Tedlar\textsuperscript{\textregistered} film. The scintillation light produced in each tile is collected by two groups of three SiPMs readout in a OR configuration, located at two opposite edges of the tile. The SiPMs are hosted on dedicated carrier boards. Each carrier board is instrumented with three SiPMs (40 $\mu m$ cell size), each with an active area of $3 \times3$ mm$^2$, made by Fondazione Bruno Kessler (FBK)~\cite{Altamura:2021wdu}.

A custom front-end board amplifies and discriminates the analog signals from the SiPMs on a carrier board; a 4-fold coincidence signal (two signals from each tile) is then required in a 50 ns time window to trigger an event. The discrimination threshold corresponds to an energy deposit of about 0.75 MeV, equivalent to about 0.4 of the most probable energy loss value of a minimum ionizing particle with normal incidence.

Data are collected via a digital signal to a GPIO pin of a Raspberry Pi 4 (RPi4)~\cite{rpi4}. 
A DT5485P CAEN module~\cite{dt5485p} is used to bias the SiPMs with a temperature compensation feedback.
The time stamp and detector position are obtained through a GPS board which uses an L80-M39 module~\cite{L80M39} connected to the RPi4. Environmental parameters (temperature, pressure and inclination/orientation) are measured by means of dedicated sensors and are recorded every 30 seconds, to be used later in the data analysis. A DS18B20 1-Wire~\cite{ds18b20} digital thermometer sensor connected to the RPi4 is used to monitor the temperature inside the detector case. A Sense HAT board, developed for the Astro Pi mission~\cite{astropi}, is used to monitor the orientation (yaw, pitch, roll) via an accelerometer, a 3D gyroscope, and a magnetometer with a LSM9DS1 sensor~\cite{lsm9ds1} for the geomagnetic field. The pressure is measured with a LPS25H sensor~\cite{lps25h} and the humidity with the HTS221 sensor~\cite{hts221}. Finally, for redundancy, a BME280 sensor~\cite{bme280,bme280_1} is also used to monitor the pressure, the temperature and the humidity inside the case.

\begin{figure}[!t]
    \centering
    \includegraphics[width=0.5\textwidth,height=6cm]{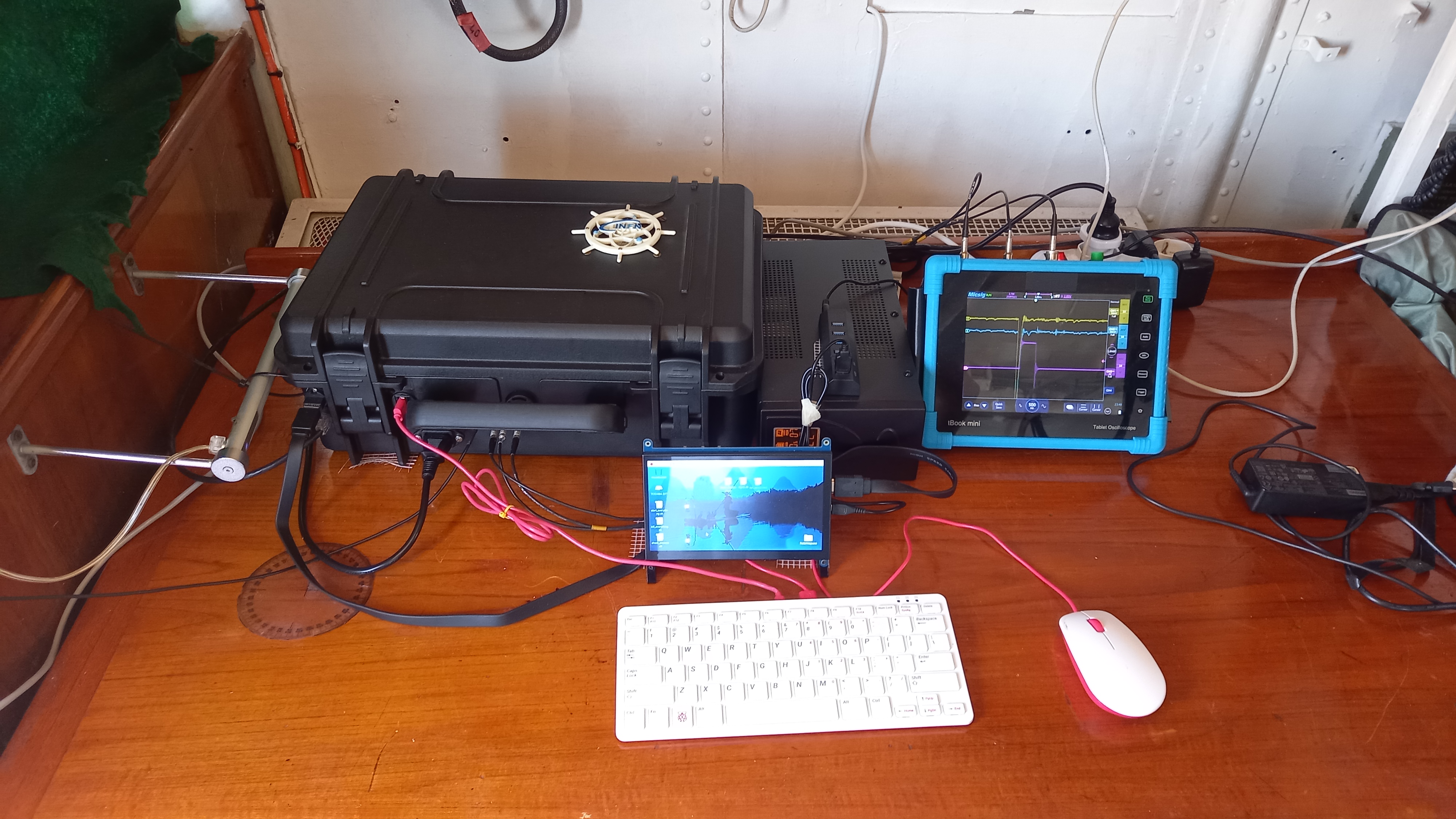}
    \includegraphics[width=0.45\textwidth,height=6cm]{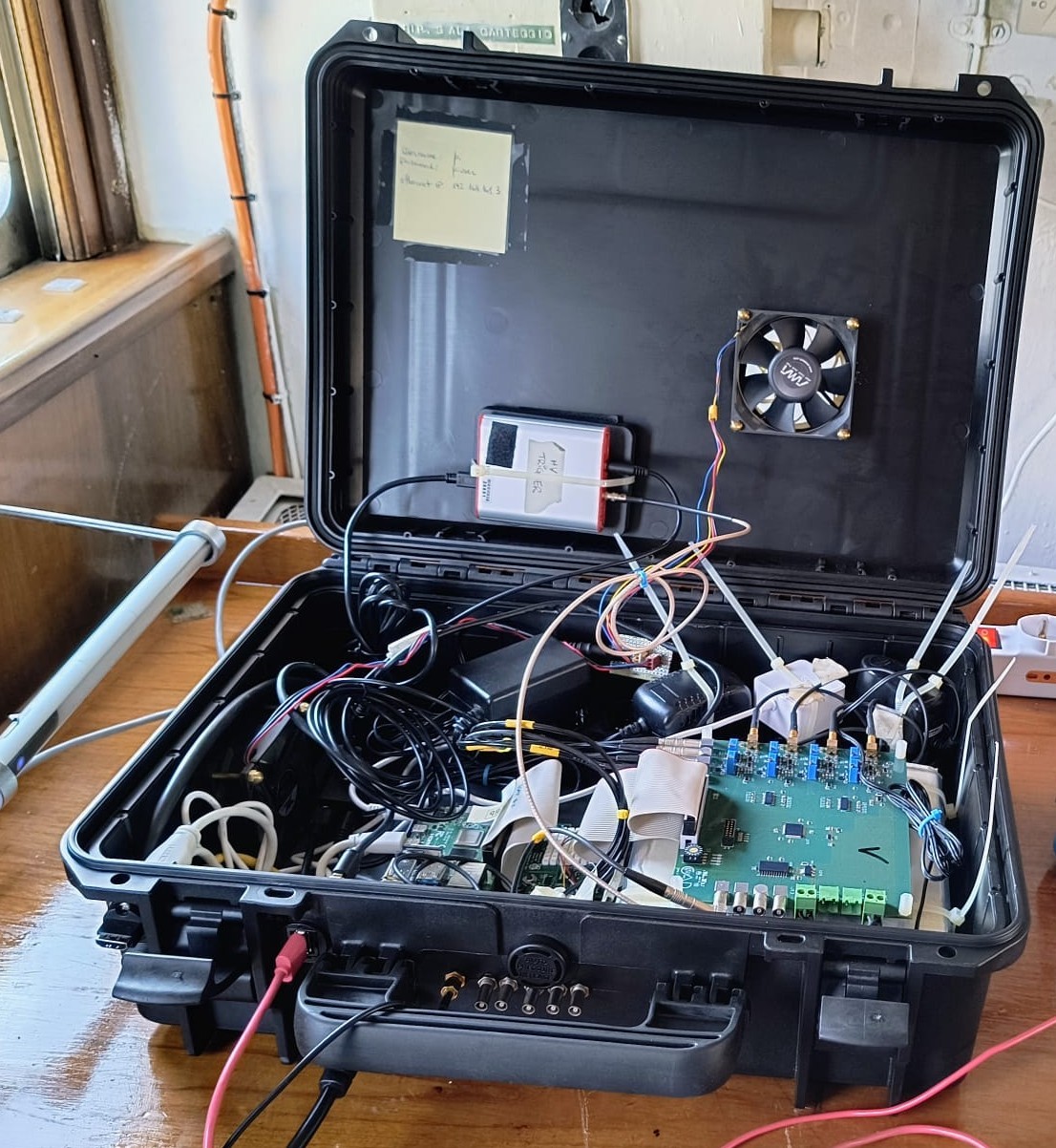}
    \caption{Left panel: the detector box installed on a desk in the Vespucci nautical room. Right panel: detector box inside view with electronic board, the SiPM bias module and the RPi4. The scintillator tiles are below the electronic board on the right side of the detector case.}
    \label{fig:det}
\end{figure}

Custom Python scripts running on the RPi4 are used to configure and readout the front-end board and the sensors. Several python scripts are also used to configure the front-end board and monitor in real time the data with the RPi4. 
The data taking was performed in short runs of 10 minutes each. The scintillator tiles, the RPi4, the DT5485P, the hard-disk and the environmental sensors are located in a black plastic case (see Fig.~\ref{fig:det}). The data are stored on a local hard-disk and sent to the INFN Bari computer center via internet connection to be stored, reconstructed and analyzed. 

\subsection*{Temperature, pressure and geomagnetic field}
\label{secB:det}

\begin{figure}[!ht]
\centering
\includegraphics[width=\textwidth]{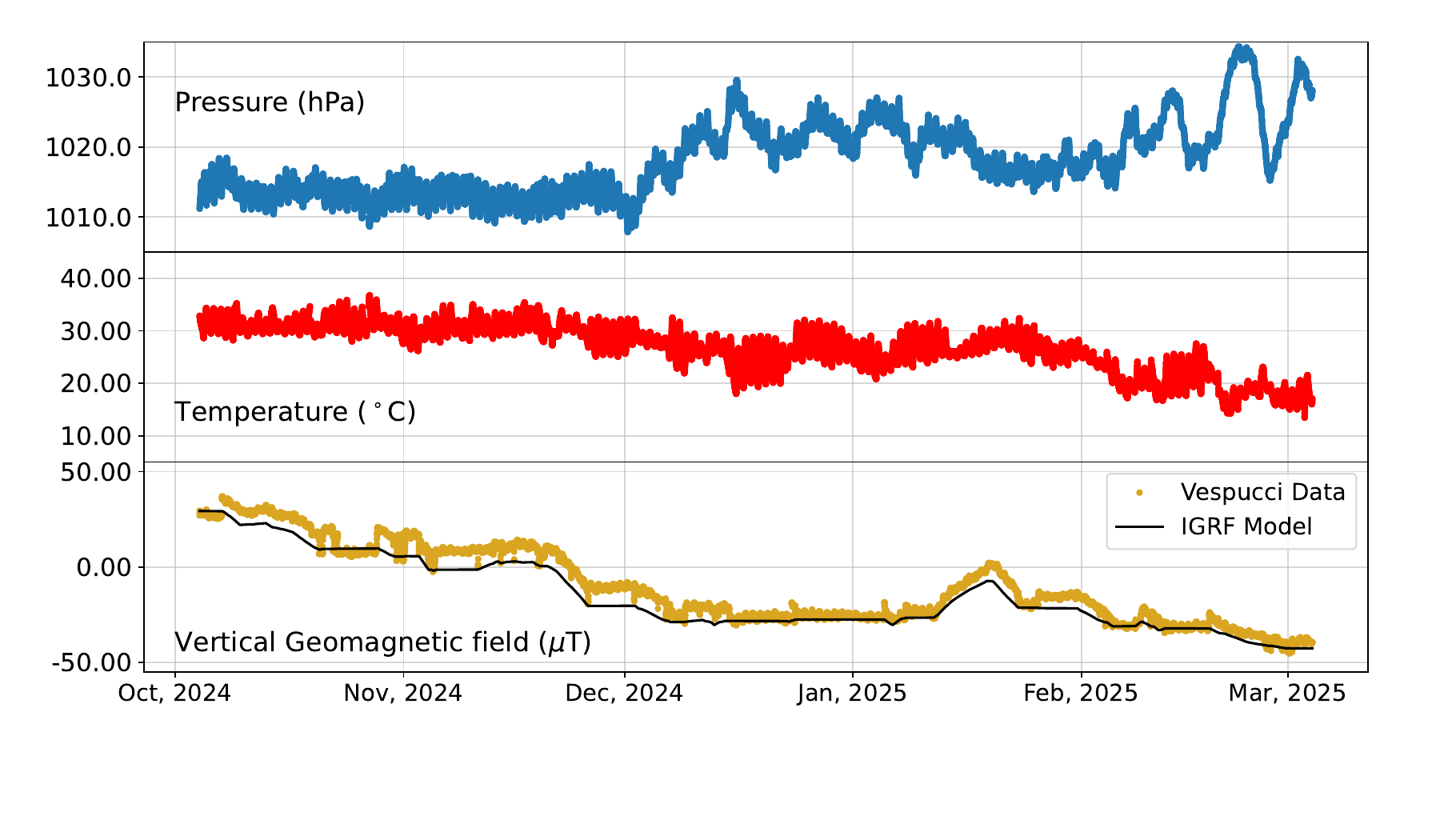}
\caption{Pressure, temperature and vertical geomagnetic field component values as function of time (UTC). In the bottom panel we also show the vertical geomagnetic field component calculated with the IGRF model. Each point refers to a single 10-min data run.}
\label{fig:TPA}
\end{figure}

In Fig.~\ref{fig:TPA} we show the temperature, the pressure and the vertical component of the geomagnetic field measured by the sensors on the detector box along the route of the Vespucci from Darwin to Trieste. The values reported in the plots of Fig.~\ref{fig:TPA} are averaged in 5 min time intervals. 

The temperature was of about 30$^\circ$ C until the end of Nov 2024, then it decreased to around 25$^\circ$ C, and from Feb 2025 it further decreased up to about 15$^\circ$ C. 
On the other hand, the pressure was approximately constant, with a value of about 1012 hPa until the end of Nov 2024; it slightly increased in Dec 2024 $ - $ Jan 2025, with a peak value of about 1030 hPa, and then we observed peak-to-peak fluctuations in the range 1015-1035 hPa. 
The orientation with respect to the vertical direction of the scintillator tiles was pretty constant for the whole data taking period.

Finally, in the bottom panel of Fig.~\ref{fig:TPA} we show the measured vertical component of the geomagnetic field by using the Sense HAT board overlaid with the one predicted by the IGRF model. We find a good agreement in the shape with a slight offset, probably due to a miscalibration of the Sense HAT magnetometer sensors or to an interference effect with other instrumentation near the detector location. 

\subsection*{Rate corrections}

The changes in the inclination of the tiles with respect to the vertical direction, due to the motion of the vessel, need to be taken into account when evaluating the count rate. 

The detector inclination affects the measured rate since its angular acceptance depends on the off-axis angle (the maximum acceptance is for the vertical direction). The rate has therefore been corrected for the inclination effect by dividing the measured values for the cosine of the average off-axis angle in each run. We found that the off-axis angles measured along the whole trip do not exceed a few degrees, and this correction has a small impact on the measured rate. 

A further correction to the measured rate is related to the variations of atmospheric pressure. An increase in pressure yields a reduction in the observed rate, since the atmospheric pressure reflects the thickness of the column of air above the vessel. This parameter determines the height where the interactions of CRs impinging on the atmosphere occur, and the amount of material that secondary particles produced by these interactions have to traverse before reaching the sea level. This effect is well known as ``barometric effect'' (see for example~\cite{Abbrescia:2020oew}).

The barometric correction to the measured rate $R_{meas}$ is evaluated as: 

\begin{equation}
    R = R_{meas} e^{ \beta (P-P_{ref})}
    \label{eq:baro}
\end{equation}
where $R$ is corrected rate, $P$ is the actual pressure, $P_{ref}$ is the average  pressure and $\beta$ is the so called ``barometric coefficient'' that was evaluated in Darwin  before the departure of the Vespucci and before the first Forbush decrease was observed in early October 2024. 
The barometric coefficient was evaluated by fitting the observed rate with the function $R = C e^{ - \beta (P-P_{ref})}$ yielding a best fit value of $\beta = (1.52 \pm 0.48) \times 10^{-3}$ hPa$^{-1}$ at $P_{ref}=1015$ hPa.
From our data we found the value $\beta = (1.52 \pm 0.48) \times 10^{-3}$ hPa$^{-1}$ at $P_{ref}=1015$ hPa.
We also re-evaluated the barometric coefficient in Trieste, and we found no significant differences 
with respect to the initial value. 
Indeed, the pressure along the path of the Vespucci vessel during its tour was quite stable, except for the last part of the journey, where we observed peak-to-peak variations up to about 15 hPa.  

No further corrections, such as those due to the thickness of materials above the detector or its detection efficiency, have been applied to the data.

\bibliographystyle{unsrt}
\bibliography{biblio.bib}

\section*{Acknowledgments}
This work was supported by an agreement between the INFN and the Italian Navy.
The authors are very grateful to the Commander ``Capitano di Vascello'' Giuseppe Lai, to the officers and to the crew of the Italian Navy tall ship Amerigo Vespucci. We extend our thanks to ``Sottocapo Aiutante'' Daniele Finocchiaro and Simone Zanni for their technical support.

The authors would like to thank the INFN Bari staff for its contribution to the procurement and to the construction of the prototype. In particular, we thank D. Dell'Olio, M. Franco, N. Lacalamita, F. Maiorano, S. Martiradonna, M. Mongelli, R. Triggiani and N. M. Aprile Ximenes for their technical support.

The authors also acknowledge the INFN Communications Office for their support to this project. In particular we thank A. Varaschin, F. Mazzotta and M. Galli for their valuable support in disseminating our activities through social media channels.

We acknowledge the NMDB database (www.nmdb.eu) founded under the European Union's FP7 programme (contract no. 213 007), and the PIs of individual neutron monitors at: Oulu (Sodankyla Geophysical Observatory of the University of Oulu, Finland).

\section*{Funding}
The authors declare that no funds or grants were received from any agency to carry out this research work.

\section*{Author contributions}
M.N. Mazziotta lead the project, wrote the manuscript, and lead the data analysis and the discussion.
F. Licciulli developed the electronic readout system.
D. Serini operated the instrument onboard the Vespucci during the route from Darwin to Singapore.
M.N. Mazziotta, F. Gargano, L. Di Venere and D. Serini installed the instrument on the Vespucci.
R. Pillera operated the ``ALBERT'' cosmic-ray detector installed at INFN Bari.
D.C., F.C., G.D.P., R.D.T., L.D.V., F.G., M.G., F.L., A.L., P.L., F.L., L.L., M.N.M., G.P., R.P. and D.S. contributed to the detector assembly, to the data analysis and participated to the discussion.

\section*{Competing interests}
The authors declare no competing interests.

\section*{Data availability}
The datasets generated during and/or analyzed during the current study are available from the corresponding author upon a reasonable request. 
 
\section*{Code availability}
Data analysis was performed using custom codes    based on software publicly available as Python~\cite{van1995python,Hunter:2007} and ROOT toolkit~\cite{Brun:1997pa}.

\section*{Additional information}

\noindent \textbf{Correspondence} and requests for materials should be addressed to M. N. Mazziotta.

\end{document}